\documentclass[12pt]{article}
\usepackage{graphicx}

\newcommand\pubnumber{NuPhys2018-Vandierendonck}
\newcommand\pubdate{\today}

\def\napoli{Department of Physics and Astronomy\\
Universiteit Gent, Belgium}
\def\support{\footnote{for the SoLid Collaboration}}

\def\Title#1{\begin{center} {\Large #1 } \end{center}}
\def\Author#1{\begin{center}{ \sc #1} \end{center}}
\def\Address#1{\begin{center}{ \it #1} \end{center}}

\newcommand\pubblock{\rightline{\begin{tabular}{l} \pubnumber\\
         \pubdate  \end{tabular}}}
\newenvironment{Abstract}{\begin{quotation}  }{\end{quotation}}
\newenvironment{Presented}{\begin{quotation} \begin{center} 
             PRESENTED AT\end{center}\bigskip 
      \begin{center}\begin{large}}{\end{large}\end{center} \end{quotation}}

\usepackage{subfigure}
\usepackage{hyperref}
\usepackage{caption}
\usepackage{lineno}

\def\beq{\begin{equation}}
\def\eeq#1{\label{#1}\end{equation}}
\def\eeqn{\end{equation}}

\def\beqa{\begin{eqnarray}}
\def\eeqa#1{\label{#1}\end{eqnarray}}
\def\eeqan{\end{eqnarray}}

\let\bar=\overbar

\def\Dslash{\not{\hbox{\kern-4pt $D$}}}
\def\dslash{\not{\hbox{\kern-2pt $\del$}}}

\def\msb{{\bar{\ssstyle M \kern -1pt S}}}

\begin{document}
\begin{titlepage}
\pubblock

\vfill
\Title{Commissioning the SoLid Detector Using Cosmic Ray Muons}
\vfill
\Author{ Giel Vandierendonck\support}
\Address{\napoli}
\vfill
\begin{Abstract}
A study of reconstructed cosmic ray muons for commissioning purposes of the SoLid detector. 
\end{Abstract}
\vfill
\begin{Presented}
NuPhys2018, Prospects in Neutrino Physics\\
Cavendish Conference Centre, London, UK, December 19--21, 2018
\end{Presented}
\vfill
\end{titlepage}
\def\thefootnote{\fnsymbol{footnote}}
\setcounter{footnote}{0}
%



\section{The SoLid Detector}

\begin{figure}[htb]
\centering
\subfigure[Four cubes with the fibres colored in green. Below, two example waveforms for scintilator signals in PVT and ZnS.]{%
\label{fig:firstSketch}%
\includegraphics[width=.25\textwidth]{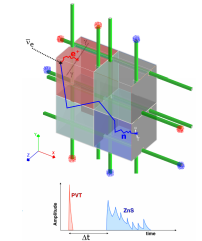}}%
\qquad
\subfigure[Components of a detector plane]{%
\label{fig:secondSketch}%
\includegraphics[width=.25\textwidth]{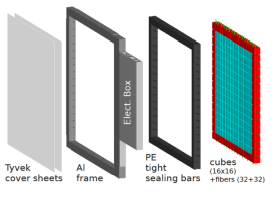}}%
\qquad
\subfigure[Schematic overview of the detector and the container in blue]{%
\label{fig:thirdSketch}%
\includegraphics[width=.25\textwidth]{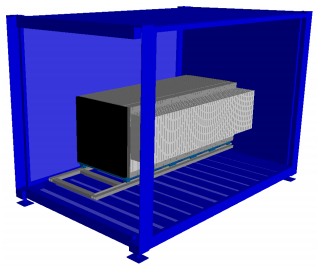}}%
\caption{Different components of the SoLid detector}
\end{figure}

The aim of SoLid~\cite{solid} is to have a measurement of the neutrino oscillations at a short baseline ([5-10] m) and to provide a measurement of the $^{235}$U energy spectrum. The SoLid detector is designed to detect electron antineutrinos using the same principle Cowan and Reines~\cite{firstneutrino} used for the first detection ever of the neutrino. The detection of the neutrino relies on identifying the reaction products of its interaction with a proton through the inverse beta decay (IBD):
\begin{equation}
\bar{\nu}_e + p \rightarrow e^+ + n
\end{equation}
\indent The protons that are used as target for the IBD interaction are primarily the nuclei of hydrogen atoms in poly-vinyl toluene (PVT) cubes that build up the SoLid detector. Furthermore, PVT is a plastic scintillator that allows for detection of the positron, which is one of the reaction products of the IBD. The other reaction product, the neutron, is first thermalised by the PVT before it will be captured on a $^6$Li nucleus in one of the additional screens of inorganic scintillator ($^6$LiF:ZnS(Ag)) that are placed on two faces of each cube. The cubes and the screens are additionally wrapped in tyvek, which is a reflective woven polymer fabric. The tyvek is not only used to keep the cube and the two screens together, but also to confine the produced scintillation light to a single cube. The light in a cube is guided to the outside of the detector volume using wavelength shifting fibres (WSFs), of which four penetrate each cube. A sketch of the cubes used in the SoLid detector can be found in Figure \ref{fig:firstSketch}.

\indent All cubes are stacked in arrays of 16 by 16 cubes in so-called detector planes. One plane is equipped by a total of 64 WSFs which are each connected to a silicon photomultiplier (SiPM) on one side and a mirror on the other side. Finally, the plane is subsequently surrounded by a polyethylene encasing, an aluminum frame and two additional large sheets of tyvek. A sketch of the detector planes can be found in Figure \ref{fig:secondSketch}\\
\indent The complete SoLid detector consists out of 5 modules of 10 planes each. The detector is placed in a container which is cooled to a temperature of [5-10]$^\circ$C. The cooled environment increases both stability and reliability of the electronics used for the readout of the detector. A sketch of the detector and the container can be found in Figure \ref{fig:thirdSketch}.

\section{Offline Muon Reconstruction}

\begin{figure}[htb]
\centering
\includegraphics[width=.7\textwidth]{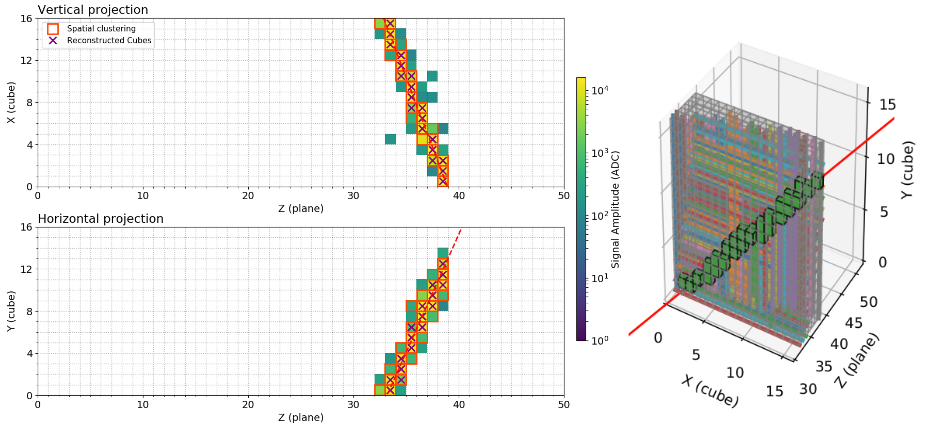}
\caption{An example of a reconstructed muon in data. (Left) seen from the fibre perspective, (right) seen from a cube perspective.}
\label{fig:reconMuon}
\end{figure}

The offline data processing of the SoLid experiment aims to distinguish between three different signals that are present in data. The first two types of signals are neutron and positron signals, which are important for the detection of IBD interactions. The third type of signals are cosmic muons, which is an important source of background.\\
\indent Muon reconstruction relies on three subsequent selections on data. First, a method called spatial clustering is applied. Spatial clustering will group all signals from neighbouring fibres and will reject all outlier fibres. This technique is useful to reject secondary particles which are produced due to spallation or ionisation by the muon. Second, an energy cut is applied on the remaining cluster of fibres. This cut is applied to reject low energetic secondary signals and light leaks. Light leaks are caused by scintillation light escaping to cubes near the cube where the interaction took place. Third, a cut on the fibre multiplicity is applied on the cluster. Muons are highly energetic and will cross a lot of cubes along their path, thus creating a cluster with a large number of fibres.\\
\indent If a cluster passed all three selections, it is tagged as a muon. Additionally, once the muon has been identified, a set of variables related to the original track of the muon are calculated in software. This set includes the polar and azimuthal angle of the muon track, its tracklength in the detector, etc ... An example of a muon track inside the SoLid detector can be found in Figure \ref{fig:reconMuon}.

\section{Muon Studies}

\begin{figure}[htb]
\centering
\subfigure[Correlation between the atmospheric pressure and the rate of reconstructed muons.]{%
\label{fig:corr}%
\includegraphics[width=.35\textwidth]{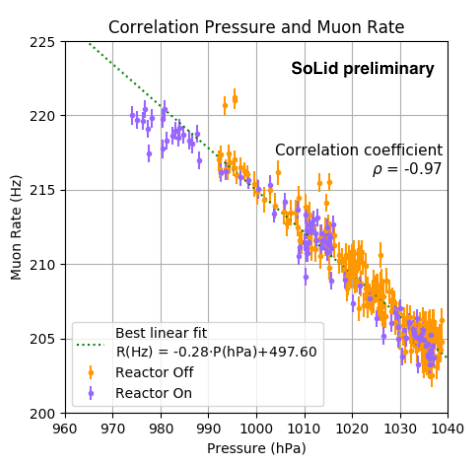}}%
\qquad
\subfigure[(Left) the distribution of time difference between consecutive energy deposits of a muon. (Right) the same distribution as a function of the plane number.]{%
\label{fig:synch}%
\includegraphics[width=.52\textwidth]{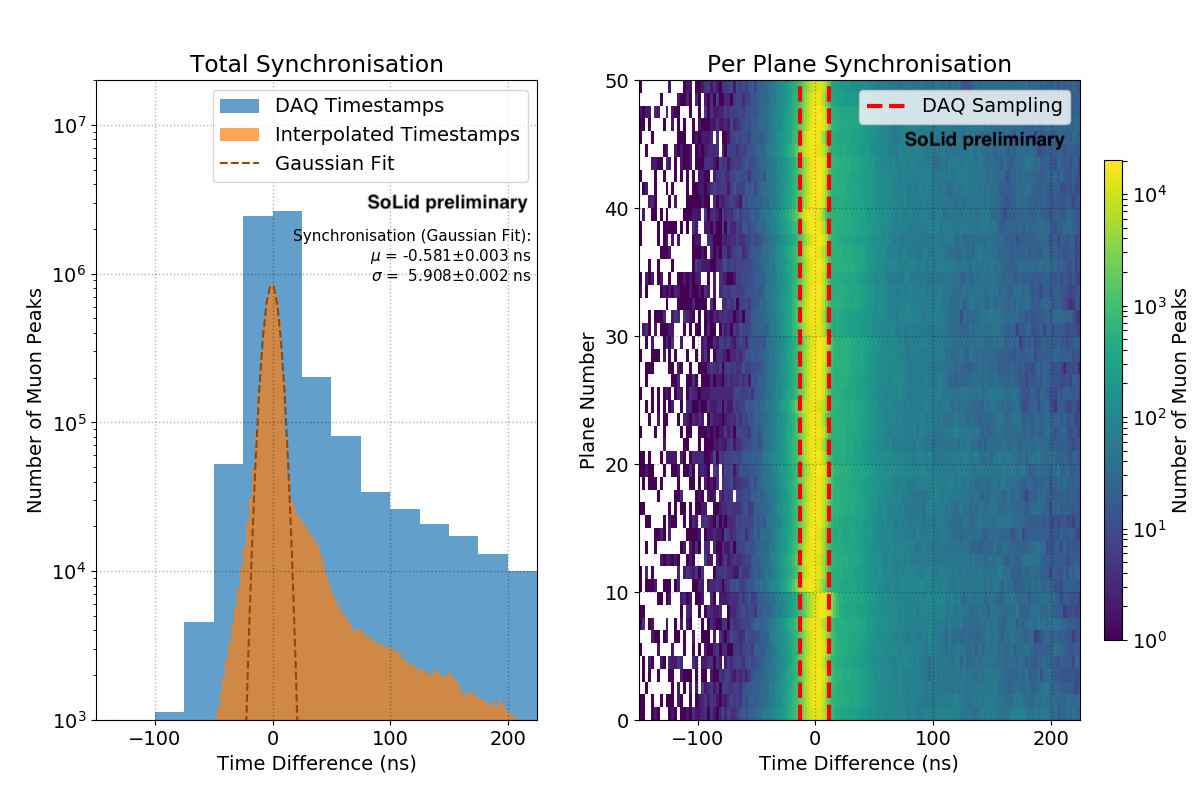}}%
\caption{Results of the muon studies.}
\end{figure}

Because the SoLid experiment has recently started taking data, a lot of verifications of the stability and the reconstruction have to be done. First of all, the stability of the data taking can be verified by monitoring the rate of muons in time. When correlating the rate of reconstructed muons with the atmospheric pressure, a linear relation is observed as can be seen in Figure~\ref{fig:corr}. As this effect has been measured before~\cite{press}, it provides confidence in the stability of the experiment.\\
\indent The time synchronisation of the detector can be studied with the reconstructed muons as well. Muons can deposit energy in a large number of planes along their path when they cross the detector. Because the time in which a muon crosses the detector is negligible compared to the time resolution of the experiment, the energy deposits are expected to be detected simultaneously. Figure \ref{fig:synch} shows the distribution of the measured time difference between energy deposits associated with a reconstructed muon. In blue, the time difference is shown in units of the DAQ timestamps (25 ns). A better time resolution was obtained by interpolating, as can be seen from the orange distribution. The detector is synchronised within [5.908 $\pm$ 0.002] ns, which is smaller than the DAQ sampling time. Figure~\ref{fig:synch} also shows the time difference as a function of the plane number to demonstrate the uniformity of the obtained time resolution across all planes.

\indent The neutron and electron reconstruction capabilities of the SoLid experiment is tested by a preliminary study of after-muon events. On one hand, stopping muons that decay inside the detector are investigated. By monitoring the time difference between a Michel electron, which is a decay product of the muon, the lifetime of the muon is calculated. A value of [2.13 $\pm$ 0.01 (stat) $\pm$ 0.07 (syst)] $\mu$s was found, which is in good agreement with the value of [2.1969811 $\pm$ 0.000002] $\mu$s that is found in literature~\cite{lifetime}. On the other hand, the neutron capture time can be calculated by investigating the time difference between a muon and a spallation neutron created by that muon. A value of [65.30 $\pm$ 1.30 (stat) $\pm$ 0.01 (syst)] $\mu$s was found, which is in good agreement with the value of [65.00 $\pm$ 0.01 (stat)] $\mu$s obtained with muons generated according to the spectrum predicted by Guan~\cite{Guan} and interfaced with a GEANT4 simulation of the detector response.

\section{Conclusions}

The SoLid detector was successfully commissioned in the winter of 2017-2018. Various tests have demonstrated that the readout and the reconstructed signals are well-understood. In particular, the time synchronisation of all detector readout signals is demonstrated using cosmic muons. In addition, the measurement of the muon decay time and the neutron capture time 
demonstrate the identification of both electron (and hence positron) and 
neutron signals with the SoLid detector. These aspects are at the basis 
to reconstruct IBD events with the SoLid experiment.

\end{document}